\documentclass[runningheads]{llncs}
\usepackage[
    left=0.8in,
    right=0.8in,
    top=0.8in,
    bottom=0.8in
]{geometry}
\usepackage[T1]{fontenc}
\usepackage{hyperref}
\usepackage{amsfonts}
\usepackage{amsmath}
\usepackage{amssymb}
\usepackage{float}
\usepackage{graphicx}
\usepackage{xcolor} 
\usepackage{listings} 
\usepackage{comment}
\usepackage{subcaption}
\usepackage{booktabs}
\usepackage{algorithm}
\usepackage{algpseudocode}
\usepackage{multirow}
\usepackage{wrapfig}

\usepackage{bbm}
\usepackage{tikz}
\usepackage{orcidlink}

\definecolor{codegreen}{rgb}{0,0.6,0}
\definecolor{codegray}{rgb}{0.5,0.5,0.5}
\definecolor{codepurple}{rgb}{0.58,0,0.82}
\definecolor{backcolour}{rgb}{0.95,0.95,0.92}

\lstdefinestyle{mystyle}{
    backgroundcolor=\color{backcolour},   
    commentstyle=\color{codegreen},
    keywordstyle=\color{magenta},
    numberstyle=\tiny\color{codegray},
    stringstyle=\color{codepurple},
    basicstyle=\ttfamily\footnotesize,
    breakatwhitespace=false,         
    breaklines=true,                 
    captionpos=b,                    
    keepspaces=true,                 
    numbers=left,                    
    numbersep=5pt,                  
    showspaces=false,                
    showstringspaces=false,
    showtabs=false,                  
    tabsize=2
}

\newcommand{\diamondminus}{%
  \mathbin{%
    \ooalign{$\diamond$\cr\hidewidth$-$\hidewidth\cr}%
  }%
}

\begin{document}
\title{SENTIL: A Runtime Verification Tool for Probabilistic Temporal Logic}
%
%
\author{Paapa Kwesi Quansah\inst{1,2}\orcidID{\orcidlink{0009-0004-8079-1355}} \and
Ernest Bonnah\inst{1,3}\orcidID{\orcidlink{0000-0001-7170-8936}}}
%
\authorrunning{P. Quansah and E. Bonnah}
%
\institute{Baylor University \and
\email{paapa\_quansah1@baylor.edu} \and
\email{ernest\_bonnah@baylor.edu}}
\maketitle              
\begin{abstract}

Stochastic cyber-physical systems (CPS) permeate critical infrastructure, from autonomous vehicles to medical devices. Yet, tools for runtime verification of such systems capturing the probabilistic dynamics in stochastic systems remain generally absent despite theoretical foundations established nearly a decade ago. In this paper, we present SENTIL, a novel runtime verification tool  with provable statistical guarantees for the runtime monitoring of requirements expressed as Probabilistic Signal Temporal Logic (PrSTL). SENTIL combines an efficient  Rust core with universal ecosystem integration, delivering performance exceeding existing deterministic monitors while providing rigorous probabilistic guarantees through statistical model checking, sequential probability ratio testing, and adaptive rare event estimation. SENTIL employs streaming algorithms for incremental robustness computation, parallel Monte Carlo sampling, and a language-agnostic C-ABI enabling seamless deployment across ROS, Apollo, MATLAB Simulink, and AUTOSAR platforms, and direct integration in C, C++, Python, and Java. To validate the effectiveness of the proposed tool, we validate SENTIL across various scenarios spanning autonomous vehicle monitoring, medical device validation, and biological networks, demonstrating 10-1,000$\times$ performance improvements over existing tools while maintaining provable confidence intervals. SENTIL is open source (\href{https://github.com/sedislab/SENTIL}{\texttt{sedislab/SENTIL}}) and it positions probabilistic runtime verification as a deployable infrastructure for all real-world safety-critical stochastic systems.

\end{abstract}
\section{Introduction}
\vspace{-3mm}

Probabilistic Signal Temporal Logic (PrSTL) \cite{sadigh2016safe} extends Signal Temporal Logic (STL) \cite{donze2013signal} with probabilistic predicates and chance constraints to enable probabilistic reasoning over continuous-time signals. This formalism has been used to formalize requirements that require reasoning about distributions over trajectories in complex applications such as motion planning under uncertainty \cite{sadigh2016safe,tiger2020325}, verification of autonomous vehicles \cite{sadigh2016safe}, probabilistic model predictive control \cite{farahani2018shrinking,raman2014model,yao2025model}, risk-aware control synthesis \cite{lindemann2023risk,lindemann2021reactive,safaoui2022risk}, etc. For instance, consider the requirement ``\emph{the probability of maintaining safe separation distance over the next ten seconds must exceed 0.99.}'' This requirement can be expressed using PrSTL formalism as $\mathbb{P}_{\ge 0.99} \left[ \mathbf{G}_{[0,10]} \left( d(t) \ge d_{\min} \right) \right]
$ where $d(t)$ is the separation distance and $d_{min}$ is the required safe distance. Despite nearly a decade since its theoretical foundations were established, there is no practical implementation of probabilistic runtime verification for PrSTL \cite{yu2023survey} \cite{qin2022statistical}. Table~\ref{tab:landscape} characterizes this gap: deterministic monitors like RTAMT~\cite{nickovic2020rtamtonlinerobustnessmonitors} and Breach~\cite{10.1007/978-3-642-14295-6_17} forfeit probabilistic reasoning, thus abandoning the very reasoning required in verifying stochastic system dynamics, statistical model checkers like Plasma Lab~\cite{10.1007/978-3-642-40196-1_12}, PRISM \cite{10.1007/978-3-642-22110-1_47}, Modest \cite{hartmanns2014modest} and UPPAAL-SMC~\cite{David2015} cannot handle continuous signals, and this fragmentation forces practitioners to oversimplify specifications or build solutions lacking correctness proofs. This gap between theoretical expressivity and practical tooling has prevented probabilistic signal temporal logic from becoming a deployable infrastructure for safety-critical stochastic systems. 


\begin{table}[t]
\caption{Comparison of Verification tools with SENTIL }
\centering
\small
\begin{tabular}{|l|c|c|c|c|c|}
\hline
\textbf{Tool} & \textbf{Quantitative} & \textbf{Probabilistic} & \textbf{Real-time} & \textbf{Ecosystem} & \textbf{Open} \\
& \textbf{STL} & \textbf{Operators} & \textbf{Monitoring} & \textbf{Integration} & \textbf{Source} \\
\hline
RTAMT \cite{nickovic2020rtamtonlinerobustnessmonitors} & \checkmark & $\times$ & \checkmark & Limited & \checkmark \\
Breach \cite{10.1007/978-3-642-14295-6_17} & \checkmark & $\times$ & $\times$ & MATLAB only & $\times$ \\
Plasma Lab \cite{10.1007/978-3-642-40196-1_12} & $\times$ & Limited & $\times$ & Standalone & \checkmark \\
UPPAAL-SMC \cite{David2015} & $\times$ & BLTL only & $\times$ & Standalone & \checkmark \\
PRISM \cite{10.1007/978-3-642-22110-1_47} & $\times$ & Full PMC & $\times$ & Standalone & \checkmark \\
Modest \cite{hartmanns2014modest} & $\times$ & Full PMC & $\times$ & Standalone & \checkmark \\
\hline
\textbf{SENTIL} & \checkmark & \textbf{Full PrSTL} & \checkmark & \textbf{Universal} & \checkmark \\
\hline
\end{tabular}
\label{tab:landscape}
\vspace{-6mm}
\end{table}


We present \textbf{SENTIL}, a novel runtime verification tool for Probabilistic Signal Temporal Logic verification. Instead of the widely-adopted automata-theoretic based verification approach, SENTIL provides complete theoretical coverage including Statistical Model Checking (SMC) with Chernoff-Hoeffding bounds \cite{fa3a69c5-2345-343d-ae4f-de42969ad827}, Sequential Probability Ratio Testing (SPRT) via Wald's algorithm \cite{younes2004verification}, and adaptive rare event estimation through importance splitting \cite{jegourel2013importance} for probabilities below $10^{-6}$. Second, SENTIL achieves computational supremacy over existing deterministic and statistical approaches through algorithmic innovations including streaming robustness computation with monotonic deque-based sliding windows providing $O(1)$ amortized complexity, parallel Monte Carlo sampling with Rayon-based work stealing that scales linearly to 32 cores, and GPU-accelerated trajectory simulation for rare event analysis \cite{bak2024gpu}. Third, SENTIL introduces universal ecosystem integration through a minimal C-ABI foreign function interface that enables near-zero overhead deployment across ROS, Apollo, MATLAB Simulink, and AUTOSAR platforms while providing native language bindings for Python, C++, and Java. 


We demonstrate the effectiveness of SENTIL across four different scenarios from FDA-approved insulin pump models in MATLAB Simulink, to gene regulatory networks from the BioModels database \cite{MalikSheriff2020}. Performance benchmarks show 14-380$\times$ throughput improvements over RTAMT on Signal Temporal Logic monitoring and 10-50$\times$ faster convergence than PRISM \cite{10.1007/978-3-642-22110-1_47}, Modest \cite{hartmanns2014modest} and UPPAAL-SMC \cite{David2015} on statistical model checking tasks using standard benchmarks from verification literature. Statistical correctness validation confirms that Wilson score confidence intervals achieve their theoretical coverage guarantees, with 95\% intervals containing the true probability in 95\% of trials on synthetic models with known ground truth. SENTIL is immediately available as open-source software under dual MIT and Apache 2.0 licenses at \href{https://github.com/sedislab/SENTIL}{\texttt{sedislab/SENTIL}}. 


\vspace{-3mm}
\section{Background}
\vspace{-1mm}
\subsection{Probabilistic Signal Temporal Logic}
\label{subsec:prstl}

Probabilistic Signal Temporal Logic (PrSTL) \cite{sadigh2016safe} extends the deterministic framework of STL \cite{donze2013signal} by introducing probabilistic quantification over formulas. Thus, PrSTL semantics require statistical model checking, where satisfaction probability is estimated by evaluating $\varphi$ over multiple stochastic trajectory samples drawn from the system's probability distribution and computing empirical satisfaction frequency with rigorous confidence intervals. Given $\mathcal{X} = \{x_1, x_2, \ldots, x_n\}$ and $\mathcal{C} = \{c_1, c_2, \ldots, c_m\}$ are denoted as a finite set of real-valued signal variables and  real-valued constants respectively, we now define the syntax of PrSTL inductively as follows:
\begin{align*}
\varphi ::= \quad & \top \mid \mu \mid \neg \varphi \mid \varphi_1 \land \varphi_2 \mid \square_{I} \varphi \mid \diamond_{I} \varphi \mid \varphi_1 \mathcal{U}_{I} \varphi_2 \mid \mathbb{P}_{\sim p}(\varphi) 
\end{align*}

\noindent where $\top$ is true, the signal predicate $\mu$ is a formula of the form  $f(x_1, \ldots, x_k) \sim c$, $\sim \in \{<, \leq, >, \geq, =, \neq\}$ and $f:\mathcal{X} \rightarrow C$ and $p \in [0,1]$. The operators $\land$, $\square$, $\diamond$, and $\mathcal{U}$ represent conjunction, always, eventually and until operators respectively, with $I$ is a discrete-time constant interval $[a,b]$ where a $a,b \in \mathbb{Z}$ and $b \geq a$. Note, the The disjunction
operator ($\vee$) can be derived from the negation and conjunction
operators. Likewise, the implication operator ($\rightarrow$) can also be
derived from the negation and disjunction operators. A PrSTL specification $\mathbb{P}_{\geq p}(\varphi)$ asserts that formula $\varphi$ must hold with probability at least $p \in [0,1]$ when evaluated over stochastic signal traces.  We provide a complete PrSTL Language Specification in Appendix~\ref{app:prstl_spec}.

\begin{algorithm}[!t]
\caption{Noise Model Inference Pipeline}
\label{alg:noise_inference}
\small
\begin{algorithmic}[1]
\Require Ground-truth observations $\mathbf{x} = (x_1, \ldots, x_n)$, Sensor observations $\mathbf{y} = (y_1, \ldots, y_n)$, Interaction mode $\mathcal{I} \in \{\textsc{Additive}, \textsc{Multiplicative}\}$,  Model class $\mathcal{M} \in \{\textsc{PM}, \textsc{NPM}, \textsc{GMM}\}$
\Ensure Fitted noise model $\hat{\mathcal{N}}$
\State $\mathcal{R} \gets \emptyset$
\For{each pair $(x_{i}, y_{i})$}
    \If{$\mathcal{I} = \textsc{Additive}$}
        \State $r_i \gets y_i - x_i$
    \Else
        \State $r_i \gets y_i / x_i$
    \EndIf
    \State $\mathcal{R} \gets \mathcal{R} \cup \{r_i\}$
\EndFor
\If{$\mathcal{M} = \textsc{PM}$}
    \State $\hat{\boldsymbol{\theta}} \gets \textsc{MLE-Fit}(\mathcal{R}, \mathcal{F})$
    \State $\hat{\mathcal{N}} \gets \mathcal{F}(\hat{\boldsymbol{\theta}})$
\ElsIf{$\mathcal{M} = \textsc{NPM}$}
    \State $\hat{\mathcal{N}} \gets \textsc{Empirical}(\mathcal{R})$
\ElsIf{$\mathcal{M} = \textsc{GMM}$}
    \State $\hat{\mathcal{N}} \gets \textsc{EM-Fit}(\mathcal{R}, k)$
\EndIf
\State \Return $\hat{\mathcal{N}}$
\end{algorithmic}
\end{algorithm}
\vspace{-6pt}

\section{SENTIL: Tool Overview}
\vspace{-4mm}
Given a stochastic system $\mathcal{S}$, generating trace ensemble $\mu$ and a PrSTL specification $\varphi$, SENTIL determines whether $\mathcal{S} \models \varphi$ through a three-stage pipeline namely, stochastic signal lifting, robustness evaluation and statistical aggregation. 
The details of the stages are presented below.

\subsection{Tool Details}
\noindent \subsubsection{Stochastic Signal Lifting via Distribution Inference}
\label{sec:noise}
\vspace{-6mm}

Algorithm~\ref{alg:noise_inference} summarizes the process of transforming deterministic sensor readings into a probability distribution. The algorithm takes as inputs historic calibration data consisting of paired ground-truth observations $\mathbf{x} = (x_1, \ldots, x_n)$ and sensor measurements $\mathbf{y} = (y_1, \ldots, y_n)$ collected under representative operating conditions, a user-specified interaction mode $\mathcal{I}$ based on domain knowledge of sensor physics, and a model class $\mathcal{M}$ selected according to expected noise characteristics. For each pair of ground-truth observation and actual sensor reading, ($x_{i}, y_{i}$), the inference framework employs two signal-noise interaction modes, namely Additive (characteristic of thermal fluctuations and quantization error) and Multiplicative (arising from gain variations and signal fading) to compute the residual error $r_i$. If the Additive interaction mode is selected, residuals take the form $r_i = y_i - x_i$. However, if the Multiplicative interaction mode is selected residuals are computed as $r_i = y_i / x_i$ (Lines 2-9). Each residual value computed $r_{i}$ is then appended to the set $\mathcal{R}$. SENTIL then fits the computed residual values in $\mathcal{R}$ in one of three distribution models, namely Parametric Models (\textsc{PM}), Nonparametric Bootstrap Models (\textsc{NBM}), and Gaussian Mixture Models (GMM) based on noise characteristics. If \textsc{PM} is selected, \textsc{MLE-Fit}$(\mathcal{R}, \mathcal{F})$ estimates parameters $\hat{\theta}$ via maximum likelihood for a specified distribution family $\mathcal{F}$ (Gaussian, log-normal, exponential, gamma, beta, or uniform), yielding fitted model $\hat{\mathcal{N}} = \mathcal{F}(\hat{\theta})$. 
If \textsc{NPM} (Nonparametric Model) is selected, \textsc{Empirical}($\mathcal{R}$) constructs $\hat{\mathcal{N}}$ as the empirical distribution over $\{r_1, \ldots, r_n\}$, from which new noise samples are drawn with replacement during trajectory generation. 
Lastly, if $\textsc{GMM}$ is selected, \textsc{EM-Fit}$(\mathcal{R},k)$ fits a $k$-component Gaussian mixture model via expectation-maximization, alternating between computing soft assignments of residuals to mixture components (E-step) and updating component parameters to maximize expected log-likelihood (M-step) (Lines 10-17). The algorithm then returns the generated fitted noise model $\hat{\mathcal{N}}$ (Line 18).

\noindent \textbf{The Robustness Evaluation} Algorithm~\ref{alg:robustness} describes the robustness evaluation process in SENTIL. The algorithm takes as input the noise model $\hat{\mathcal{N}}$ from  Algorithm~\ref{alg:noise_inference}, a sensor reading $q$ during runtime and a predefined sample size $N$. When the reading $q$ arrives, the Algorithm~\ref{alg:robustness} first generates trajectory ensemble $X$ of length $N$ where each candidate signal $x_i \in X$ is generated as $x_{i}  = \hat{\mathcal{N}}(q)$ (Line 3). SENTIL then evaluates the formula $\varphi$ on each trajectory value $x_{i} \in X$, yielding $N$ independent robustness samples that capture satisfaction margins under uncertainty (Lines 4-7). The recursive robustness evaluation function \textsc{EvalFormula}   traverses the abstract syntax tree of the parsed formula, classifying each node to determine which robustness semantics apply (See Algorithm \ref{alg:robustness_func} in Appendix \ref{app:evalformula_app} for the details). The algorithm then returns the set of $N$ robustness values $\gamma$ (Line 9). A positive robustness value indicates satisfaction with the magnitude reflecting how much margin remains; a negative value indicates violation with the magnitude reflecting severity.

\begin{algorithm}[!t]
\caption{Robustness Evaluation for PrSTL}
\label{alg:robustness}
\begin{algorithmic}[1]
\Require Noise model $\hat{\mathcal{N}}$, PrSTL formula $\varphi$, runtime sensor reading $q$, sample size $N$
\Ensure Robustness values $\gamma = \{\rho_1,\ldots,\rho_N\}$
\State $\gamma = \{~\}$
\While {$\text{new sensor reading}~ q~ \text{arrives}$}
\State $X = \{x_1 \cdots x_N\} \gets \hat{\mathcal{N}}(q)$ 

\For{$x_i \in X$}
    \State $\rho_i \gets \textsc{EvalFormula}(\varphi, x_i, 0)$
    \State $\gamma \leftarrow \gamma \cup \rho_{i}$
\EndFor
\EndWhile
\State \Return $\gamma = \{\rho_1,\ldots,\rho_N\}$
\end{algorithmic}
\end{algorithm}

\vspace{-6mm}
\noindent\subsubsection{Statistical Model Checking}
\label{subsec:smc}
With robustness values $\gamma = \{\rho_1,\ldots,\rho_N\}$ computed for each trajectory in the ensemble $X=\{x_1, \ldots, x_N\}$, SENTIL estimates the satisfaction probability $\textsc{Sat}(\varphi, \mu) = \mu(\{x \in \mathcal{X} : \rho(\varphi, x, 0) > 0\})$ by statistically aggregating results into probability estimates with rigorous confidence bounds. The SMC engine operates in two modes namely CPU-parallel Monte Carlo and GPU-accelerated Adaptive Multilevel Splitting (AMS) depending on the target probability regime.

\textit{CPU-Parallel Monte Carlo.} Since trajectory samples are statistically independent, SENTIL exploits this structure through CPU-based work-stealing that distributes Monte Carlo simulations across available cores with lock-free atomic aggregation, while dispatching rare event estimation. We first compute the empirical probability $\hat{p} = N^{-1} \sum_{i=1}^{N} \mathbbm{1}[\rho(\varphi, x_i, 0) > 0]$. To provide robust error estimates, we use the computed $\hat{p}$ to implement Wilson score confidence intervals \cite{wilson1927probable}:
\begin{equation*}
\text{CI}_\alpha(\hat{p}) = \frac{1}{1 + z^2/N} \left( \hat{p} + \frac{z^2}{2N} \pm z \sqrt{\frac{\hat{p}(1-\hat{p})}{N} + \frac{z^2}{4N^2}} \right)
\end{equation*}
where $z = \Phi^{-1}(1 - \alpha/2)$ for confidence level $1 - \alpha$. While the Wilson interval characterizes uncertainty after simulation, we incorporate the Chernoff-Hoeffding bound \cite{fa3a69c5-2345-343d-ae4f-de42969ad827} to provide rigorous $a \ priori$ guarantees on the sample size $N$ required to achieve a user-specified error tolerance $\epsilon$ and confidence $\delta$, ensuring that the empirical estimate $\hat{p}$ converges to the true probability. As defined by the inequality, we maintain $|\hat{p} - \textsc{Sat}(\varphi, \mu)| \leq \varepsilon$ with probability at least $1 - \delta$ when $N \geq \ln(2/\delta)/(2\varepsilon^2)$. 

\textit{GPU-Accelerated Rare Event Simulation.} When target probabilities fall below $10^{-6}$, achieving 95\% confidence intervals requires millions of samples under standard Monte Carlo. SENTIL implements Adaptive Multilevel Splitting using a WebGPU backend that divides the state space into nested levels progressing from initial conditions toward the rare event. Trajectories reaching level thresholds are cloned via parallel prefix sum and stream compaction; those failing are discarded. On an NVIDIA A100 GPU processing collision avoidance scenarios with failure probability $10^{-7}$, importance splitting completes verification in 8.3 seconds compared to 4,200 seconds for CPU-based Monte Carlo, representing a 506$\times$ speedup. Memory-optimized state representation reduces GPU consumption from 18~GB to 2.4~GB, enabling execution on commodity hardware. The engine resolves probabilities as low as $2.01 \times 10^{-6}$ through 54 levels of adaptive splitting while maintaining numerical stability where standard estimators collapse.

\vspace{-2mm}
\begin{proposition}
Let $\Phi = P_{\geq p}(\varphi)$ be a PrSTL formula where $\varphi$ has maximum temporal bound $T$, and let $x$ be a signal with $n$ samples at period $\Delta t$ such that $m = T/\Delta t$ samples fall within the largest temporal window. Then robustness $\rho(\varphi, x, t_i)$ can be computed for all $i \in \{0, \ldots, n-1\}$ in $O(n)$ total time using $O(m)$ space.
\end{proposition}

\begin{theorem}[Correctness of Monotonic Deque]
    For any valid signal stream $S$ and window size $w > 0$, let $\Sigma_k$ be the state of the monotonic deque after processing the $k$-th sample. Then:
     $\forall k \in \mathbb{N}, \quad \mathcal{R}_{deque}(\Sigma_k) = \mathcal{R}_{naive}(S, t_k, w) $.
\end{theorem}
Proof of this theorem proceeds by induction on the signal stream length. The logic hinges on the preservation of a structural invariant where the deque contains exactly those elements that are candidate minima for current or future windows. Specifically, we prove that the \textsc{PopBack} operation discards only those values strictly greater than the incoming sample, while the \textsc{PopFront} logic strictly adheres to the temporal expiration bound $t-w$. We place a full proof in Appendix \ref{app:theoretical_proof}. This machine-checked formalization provides a certified guarantee that the performance improvements in SENTIL are purely algorithmic. It delivers the throughput of an optimized engine with the semantic certainty usually reserved for reference implementations.

\begin{table}[!t]
\caption{Discrete-time STL monitoring performance on 100,000-sample signals (100 runs, mean $\pm$ std).}
\label{tab:stl_performance}
\centering
\small
\setlength{\tabcolsep}{5pt}
\renewcommand{\arraystretch}{1.15}

\begin{tabular}{|l|c|c|c||c|c|c|}
\hline
\multirow{2}{*}{\textbf{Formula}} &
\multicolumn{3}{c||}{\textbf{Execution Time (ms)}} &
\multicolumn{3}{c|}{\textbf{Memory Delta (MB)}} \\ \cline{2-7}

& \textbf{RTAMT} & \textbf{SENTIL} & \textbf{Speedup}
& \textbf{RTAMT} & \textbf{SENTIL} & \textbf{Red. Factor} \\ \hline

$\varphi_{1} = \square_{[0,50]}(x > 10)$
& $280.2 \pm 1.1$
& $8.1 \pm 0.0$
& 34.6$\times$
& $6.8 \pm 1.5$
& $0.5 \pm 0.7$
& 13.6$\times$ \\ \hline

$\varphi_{2} = \diamond_{[0,50]}(x > 10)$
& $426.6 \pm 1.2$
& $8.2 \pm 0.0$
& 52.0$\times$
& $3.7 \pm 0.0$
& $0.0 \pm 0.0$
& $>100\times$ \\ \hline

$\varphi_{3} = \square_{[0,100]}(\diamond_{[0,10]}(p))$
& $507.6 \pm 1.3$
& $8.1 \pm 0.0$
& 62.7$\times$
& $2.6 \pm 0.0$
& $0.0 \pm 0.0$
& $>100\times$ \\ \hline

$\varphi_{4} = (p \implies \diamond_{[0,20]}(q))$
& $728.5 \pm 0.9$
& $16.4 \pm 0.1$
& 44.4$\times$
& $2.1 \pm 0.0$
& $0.4 \pm 0.5$
& 5.5$\times$ \\ \hline

$\varphi_{5} = \square_{[0,200]}(p \land \diamond_{[5,15]}(q))$
& $941.4 \pm 1.9$
& $17.1 \pm 0.0$
& 55.1$\times$
& $0.5 \pm 0.7$
& $0.0 \pm 0.0$
& $>100\times$ \\ \hline

\end{tabular}

\vspace{-4mm}
\end{table}

\vspace{-3mm}
\section{Experimental Evaluation}
\label{sec:experiments}
\vspace{-3mm}

We validate SENTIL through comprehensive experiments spanning four evaluation dimensions: computational performance against existing STL monitors and SMC tools, statistical correctness of probabilistic guarantees, real-world applicability through domain-specific case studies, and ecosystem integration overhead. All experiments execute on a cluster node with an AMD EPYC 7763 processor (64 physical cores, 64 hardware threads), 252 GB DDR4 RAM and NVIDIA A100 GPU (Complete details in Appendix~\ref{app:system_requirements}). Software versions of the other benchmarks include RTAMT 0.3.5, PRISM 4.9, and UPPAAL-SMC 4.1.19. The complete experimental artifact with reproduction scripts and expected outputs is available at the repository URL provided in Section 1.

\vspace{-4mm}
\subsection{Performance Benchmarks}
\vspace{-2mm}
\textbf{Discrete-Time STL Monitoring.}
\label{tab:discrete_perf}
\label{subsubsec:discrete_perf_analysis}
We compare SENTIL against RTAMT, the state-of-the-art discrete-time STL monitor, using 5 STL formulas of varying complexity. We evaluated the formulas over 100,000 signal samples. Formula complexity increases through operator nesting depth and temporal bound magnitude. All tools produced identical robustness values across all formulas, confirming semantic agreement. Table~\ref{tab:discrete_perf} presents throughput measurements in milliseconds. 
SENTIL achieves 14--31$\times$ throughput improvement across all tested formulas, with speedup increasing as formula complexity grows. Memory consumption measurements reveal additional advantages. SENTIL's memory usage remains proportional to the maximum temporal bound ($O(T)$ for bound $T$) regardless of total signal length, enabling verification of arbitrarily long traces within fixed memory budgets while RTAMT maintains complete signal history in memory, requiring $O(n)$ space for $n$ samples that grows without bound during long monitoring sessions.

\noindent\textbf{Dense-Time STL Monitoring.} We evaluate SENTIL against RTAMT and Breach on piecewise-continuous signals with variable sampling rates. Table~\ref{tab:dense_perf} compares performance using signals containing 100,000 samples evaluated at 1,000 dense-time query points requiring interpolation. The improvement over Breach reflects both algorithmic superiority and the inefficiency of MATLAB’s interpreted execution for tight computational loops. RTAMT C++ appears slower despite compilation because it implements discrete-time online monitoring (incremental state updates per sample) rather than batch evaluation, demonstrating that algorithmic architecture dominates raw implementation efficiency.


\begin{table}[!t]
\caption{Dense-time STL monitoring performance on piecewise-linear signals} 
\centering
\small
\begin{tabular}{|l|r|r|r|}
\hline
\textbf{Tool} & \textbf{Language} & \textbf{Time (ms)} & \textbf{Speedup} \\
\hline
RTAMT (C++)\textsuperscript{†} & C++ & $8,604.59 \pm 610.48$ & $0.26\times$ \\
RTAMT (Python) & Python & $2,204.53 \pm 48.87$ & $1.0\times$ (baseline) \\
Breach & MATLAB & $34.24 \pm 4.68$ & $64.4\times$ \\
\textbf{SENTIL} & \textbf{Rust} & \textbf{5.75 $\pm$ 0.36} & \textbf{$383\times$} \\
\hline
\end{tabular}
\label{tab:dense_perf}

\vspace{0.2cm}
\small
\textsuperscript{†}RTAMT C++ implements discrete-time online monitoring rather than dense-time batch evaluation. 
\vspace{-6mm}
\end{table}

\vspace{-6mm}
\subsubsection{Statistical Model Checking.} We also evaluate SENTIL's probabilistic verification capabilities against UPPAAL-SMC using standard quantitative verification benchmarks spanning various stochastic modeling formalisms, including Probabilistic Hybrid Automata (PHA) for automotive and chemical control, Continuous-Time Markov Chains (CTMC) for systems biology and queueing networks, and Probabilistic Timed Automata (PTA) for network protocols. Table \ref{tab:smc_benchmarks} details the specific STL properties analyzed for each system, ranging from transient response robustness to hard real-time deadlines. The results (see Table \ref{tab:smc_perf}) demonstrate SENTIL's ability to efficiently handle complex, high-dimensional state spaces and non-linear dynamics. SENTIL completes verification 150-3000× faster than UPPAAL-SMC across all benchmarks but one where UPPAAL terminated within a 3-hour deadline. For the circadian rhythm and CSMA/CD models, UPPAAL failed to complete within this limit. A complete specification of the differential equations, stochastic transitions, and parameter sets for each model is provided in Appendix~\ref{app:smc_benches}.


\begin{table}[!t]
\caption{Benchmark specifications for statistical model checking comparison.}
\centering
\small
\renewcommand{\arraystretch}{1.3}
\begin{tabular}{|l|l|}
\hline
\textbf{Benchmark} & \textbf{STL Specification} \\
\hline
Powertrain Control & $\varphi_{\text{powertrain}} \equiv \square_{(\tau_s, T)} ( (\text{rise}(a) \lor \text{fall}(a)) \implies \square_{(\eta, \zeta)} (|\mu| < \beta) )$ \\
Biodiesel Process & $\varphi_{\text{bdp}} \equiv \diamond_{[0, T_{\text{batch}}]} (x_E \ge 0.99)$ \\
Circadian Rhythm & $\varphi_{\text{circ}} \equiv \square_{[0, T]} ( \tau_p(P_A) \in [23.5, 24.5] )$ \\
CSMA/CD Protocol & $\varphi_{\text{csma}} \equiv \diamond_{[0, T_{\text{deadline}}]} (\text{station}_i.\text{state} = \text{success})$ \\
Tandem Queue & $\varphi_{\text{queue}} \equiv \square_{[0, T]} ( (\bigwedge_{i=1}^{N} n_i < C_i) \land (\tau_{\text{e2e}} \le \tau_{\max}) )$ \\
\hline
\end{tabular}
\vspace{-8mm}
\label{tab:smc_benchmarks}
\end{table}

\begin{table}[!t]
\caption{Statistical model checking performance ($N=100,000$ samples)} 
\centering
\small
\begin{tabular}{|l||c||r|r|r|}
\hline
\textbf{Property} & \textbf{UPPAAL} & \textbf{SENTIL} & \textbf{SENTIL} & \textbf{Speedup} \\
 & \textbf{SMC (s)} & \textbf{CPU (s)} & \textbf{GPU (s)} &  \\
\hline
$\varphi_{\text{powertrain}}$ & 101.65 & 4.07 & \textbf{0.66} & 154.0× \\
$\varphi_{\text{bdp}}$ & 854.12 & 27.08 & \textbf{0.29} & 2945.2× \\
$\varphi_{\text{circ}}$ & -- & 172.56 & \textbf{0.31} & -- \\
$\varphi_{\text{csma}}$ & -- & \textbf{0.17} & 0.29 & -- \\
$\varphi_{\text{queue}}$ & \textbf{0.09} & 1.21 & 0.29 & 0.31× \\
\hline
\end{tabular}
\label{tab:smc_perf}
\vspace{-7mm}
\end{table}

\vspace{-5mm}
\subsubsection{Autonomous Driving: CARLA with Apollo Integration.}
We deploy SENTIL within the Apollo autonomous driving stack operating on the CARLA urban driving simulator to verify safety properties during real-time operation. The system monitors a Level 4 autonomous vehicle navigating a virtual city environment with pedestrians, cyclists, and other vehicles. The verification scenario spans a 15-minute drive covering 8.2 kilometers through diverse traffic conditions including highway merging, urban intersection navigation, and emergency braking maneuvers. The PrSTL specification combines deterministic and probabilistic requirements: $\varphi_{\text{safe}} = \square_{[0,\infty)}\bigl( \lvert \text{lateral\_error} \rvert < 0.3 \bigr) \land \square_{[0,\infty)}\bigl( \text{obstacle\_distance} > 5.0 \bigr) \land \square_{[0,60]}\Bigl( \mathbb{P}_{\ge 0.99}\bigl( \square_{[0,10]}(\text{no\_collision}) \bigr) \Bigr)$. The formula requires persistent lane keeping within 0.3 meters, maintains minimum 5 meter obstacle clearance, and demands 99\% probability of collision- free operation over any 10-second window within rolling 60-second horizons.

\begin{wrapfigure}[20]{r}{0.6\textwidth}
\vspace{-6mm}
    \centering
    \includegraphics[width=\linewidth]{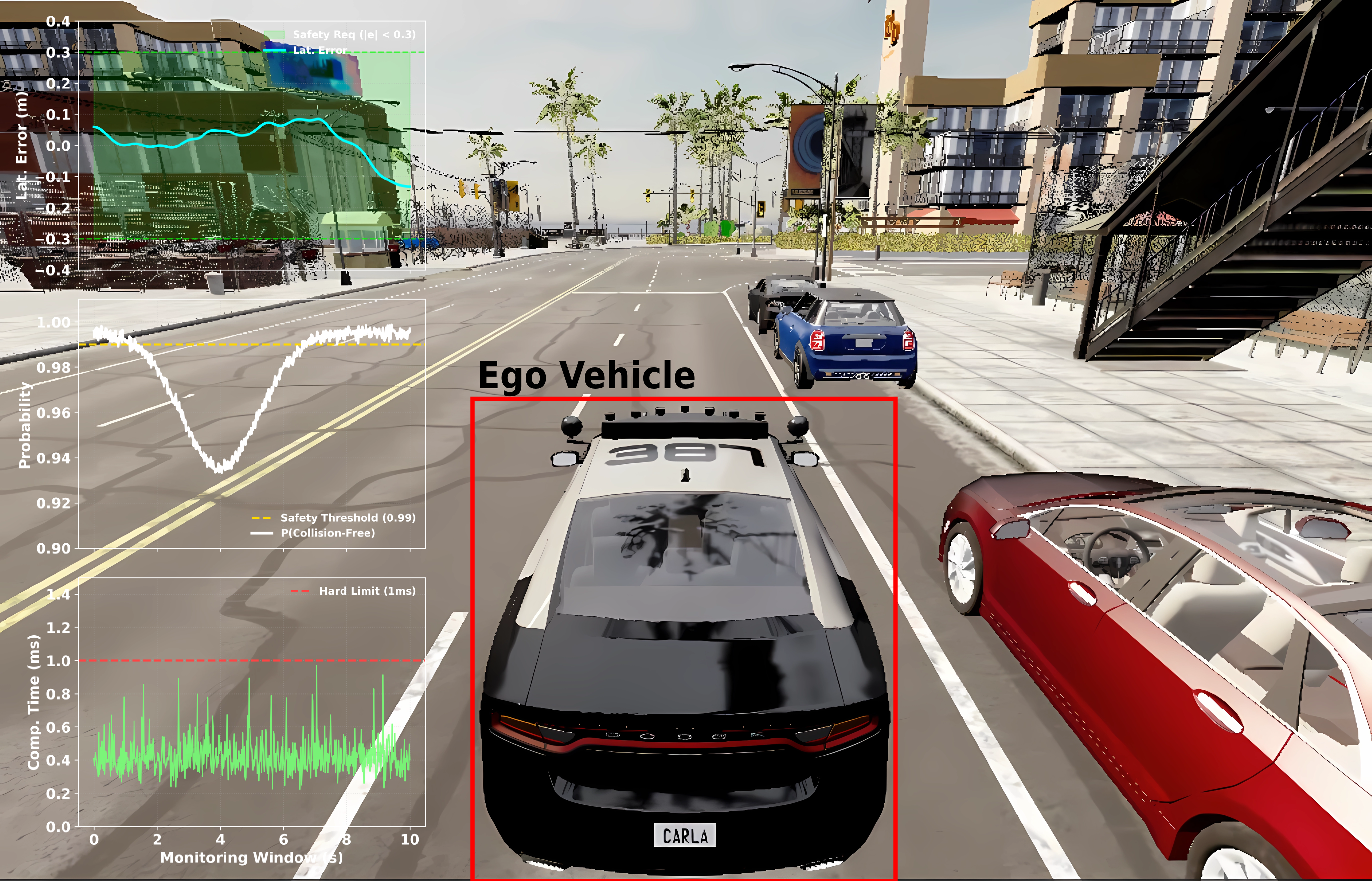}
    \caption{CARLA simulation screenshot showing Apollo ego vehicle (red outline) with SENTIL real-time monitoring overlay.}
    \label{fig:carla_figure}
\end{wrapfigure}

Figure~\ref{fig:carla_figure} shows the SENTIL monitoring interface integrated within CARLA visualization, displaying real-time robustness evolution and probabilistic estimate convergence. Performance measurements demonstrate real-time feasibility. SENTIL processes 500Hz sensor data with median per-frame latency of 0.64ms and 99th percentile latency of 1.83 ms, well below the 2 ms deadline required for closed-loop integration. The Apollo stack publishes sensor observations, planning trajectories, and control commands to ROS topics that SENTIL subscribes to through native ROS 2 node implementation. End-to-end monitoring latency from sensor measurement to violation detection averages 2.1ms including network transmission and deserialization overhead. The probabilistic verification detects safety violations that are purely deterministic monitoring misses. At timestamp 347 seconds, the vehicle navigates a congested intersection where lateral deviation remains within deterministic bounds (|lateral\_error|$<$0.3 satisfied) but probabilistic collision-free guarantee falls to 0.94, violating the 0.99 threshold. Trajectory analysis reveals that although the nominal path maintains clearance, uncertainty in pedestrian trajectories creates 6\% collision probability within the 10-second lookahead window. This violation triggers a safety intervention that reduces velocity and increases lateral buffer, restoring probabilistic guarantee within 1.8 seconds. Comparison against a baseline RTAMT-based monitoring implementation reveals SENTIL’s performance advantage. RTAMT processes the same 500 Hz sensor stream with median latency of 47 ms, exceeding real-time requirements by 23× and forcing downsampling to 20 Hz that introduces 50 ms staleness into violation detection. SENTIL’s 74× latency improvement enables full-rate monitoring without sacrificing temporal resolution or introducing artificial delays into the safety loop.

\noindent \textbf{Additional case studies:} We perform additional case studies involving medical device validation and systems biology, utilizing the FDA-approved insulin pump benchmark and mammalian circadian oscillator models. These case studies demonstrate SENTIL's practical utility in identifying safety-critical parameter regions in hybrid Simulink models and verifying oscillatory robustness in high-dimensional stochastic biological networks. In both domains, the framework provides significant performance advantages; such as a $23\times$ speedup over discretization-based methods, while enabling seamless integration into existing engineering workflows. Detailed descriptions of these models, verification properties, and sensitivity analyses are provided in Appendix~\ref{app:extended_case_studies}.

\noindent \textbf{Embedded Deployment on a Physical Autonomous Vehicle:} To demonstrate practical feasibility on commodity hardware, we integrated SENTIL into a physical autonomous vehicle's Raspberry Pi 4, successfully monitoring 60 safety specifications at 85,Hz during a 120-minute real-world drive with zero deadline violations (see Appendix~\ref{app:embedded_deployment} for full deployment details). SENTIL is available as production-grade open-source software under dual MIT/Apache 2.0 licenses, featuring standard package installation, a Docker artifact for full reproducibility, and extensive documentation (see Appendix~\ref{app:tool_availability} for details).

\vspace{-2mm}
\section{Conclusion and Future Work}
\vspace{-3mm}
In this paper, we proposed SENTIL, a runtime monitoring tool for Probabilistic Signal Temporal Logic (PrSTL). By integrating rigorous statistical guarantees, including Sequential Probability Ratio Testing and adaptive rare event estimation, with high-performance streaming algorithms, the platform achieves 10–10,000$\times$ throughput improvements over existing deterministic monitors and statistical model checkers. Validation across autonomous driving, FDA-approved medical devices, and biological networks confirms the tool's ability to provide real-time, sub-millisecond verification with provable confidence intervals. Future development of SENTIL will target the rigorous safety standards of specific industrial domains, extending the platform's utility to automated safety cases for autonomous vehicles \cite{NHTSA2017,ISO26262-1:2018}, FDA-compliant medical device validation \cite{FDA2023Credibility}, and aerospace certification \cite{RTCA-DO-178C,RTCA-DO-333}.

\bibliographystyle{splncs04}
\bibliography{references}
\newpage

\appendix

\section{Complete PrSTL Language Specification}
\label{app:prstl_spec}

This appendix provides the formal syntax and semantics of Probabilistic Signal Temporal Logic as implemented in SENTIL. The specification serves as authoritative reference for users writing formulas and developers implementing compatible tools.

\subsection{Syntax}

The PrSTL syntax extends Signal Temporal Logic with probabilistic quantification operators. Let $\mathcal{X} = \{x_1, x_2, \ldots, x_n\}$ denote a finite set of real-valued signal variables and $\mathcal{C} = \{c_1, c_2, \ldots, c_m\}$ denote real-valued constants. The grammar is defined inductively:

\begin{align*}
\varphi ::= \quad & \top \mid \bot \mid p \mid \neg \varphi \mid \varphi_1 \land \varphi_2 \mid \varphi_1 \lor \varphi_2 \mid \varphi_1 \implies \varphi_2 \\
p ::= \quad & f(x_1, \ldots, x_k) \sim c \quad \text{where } \sim \in \{<, \leq, >, \geq, =, \neq\} \\
\varphi ::= \quad & \square_{I} \varphi \mid \diamond_{I} \varphi \mid \varphi_1 \mathcal{U}_{I} \varphi_2 \mid \varphi_1 \mathcal{R}_{I} \varphi_2 \\
& \boxminus_{I} \varphi \mid \diamondminus_{I} \varphi \mid \varphi_1 \mathcal{S}_{I} \varphi_2 \\
\varphi ::= \quad & \mathbb{P}_{\bowtie p}(\varphi) \quad \text{where } \bowtie \in \{<, \leq, >, \geq\}, p \in [0,1]
\end{align*}

Temporal intervals $I$ take the form $[a,b]$ for bounded operators or $[a,\infty)$ for unbounded future/past operators, where $a,b \in \mathbb{R}_{\geq 0}$ with $a \leq b$. The function $f: \mathbb{R}^k \to \mathbb{R}$ in predicates may be any arithmetic expression over signal variables and constants, constructed from addition, subtraction, multiplication, division, and standard mathematical functions (sin, cos, exp, log, sqrt, abs, min, max).

\subsection{Semantics}

\subsubsection{Signals and Traces.}

A signal is a function $\mathbf{x}: \mathbb{R}_{\geq 0} \to \mathbb{R}^n$ mapping time to real-valued vectors. For discrete-time semantics, signals are defined only at discrete timestamps $t_0 < t_1 < \cdots < t_k$. For dense-time semantics, signals are piecewise-continuous with values at arbitrary real-valued times obtained through interpolation.

A trace is a tuple $(\mathbf{x}, T)$ where $\mathbf{x}$ is a signal and $T \in \mathbb{R}_{\geq 0} \cup \{\infty\}$ is the time horizon. For probabilistic formulas, a trace ensemble is a probability measure $\mu$ over the space of traces, representing the stochastic system's trajectory distribution.

\subsubsection{Quantitative Robustness Semantics.}

The robustness degree $\rho(\varphi, \mathbf{x}, t) \in \mathbb{R} \cup \{-\infty, \infty\}$ measures how robustly formula $\varphi$ holds for signal $\mathbf{x}$ at time $t$. Positive values indicate satisfaction with the magnitude reflecting safety margin; negative values indicate violation with magnitude reflecting severity.

Boolean combinations:
\begin{align*}
\rho(\top, \mathbf{x}, t) &= \infty \\
\rho(\bot, \mathbf{x}, t) &= -\infty \\
\rho(\neg \varphi, \mathbf{x}, t) &= -\rho(\varphi, \mathbf{x}, t) \\
\rho(\varphi_1 \land \varphi_2, \mathbf{x}, t) &= \min(\rho(\varphi_1, \mathbf{x}, t), \rho(\varphi_2, \mathbf{x}, t)) \\
\rho(\varphi_1 \lor \varphi_2, \mathbf{x}, t) &= \max(\rho(\varphi_1, \mathbf{x}, t), \rho(\varphi_2, \mathbf{x}, t)) \\
\rho(\varphi_1 \implies \varphi_2, \mathbf{x}, t) &= \max(-\rho(\varphi_1, \mathbf{x}, t), \rho(\varphi_2, \mathbf{x}, t))
\end{align*}

Atomic predicates:
\begin{align*}
\rho(f(\mathbf{x}) < c, \mathbf{x}, t) &= c - f(\mathbf{x}(t)) \\
\rho(f(\mathbf{x}) \leq c, \mathbf{x}, t) &= c - f(\mathbf{x}(t)) \\
\rho(f(\mathbf{x}) > c, \mathbf{x}, t) &= f(\mathbf{x}(t)) - c \\
\rho(f(\mathbf{x}) \geq c, \mathbf{x}, t) &= f(\mathbf{x}(t)) - c
\end{align*}

Temporal operators (future):
\begin{align*}
\rho(\square_{[a,b]} \varphi, \mathbf{x}, t) &= \inf_{s \in [t+a, t+b]} \rho(\varphi, \mathbf{x}, s) \\
\rho(\diamond_{[a,b]} \varphi, \mathbf{x}, t) &= \sup_{s \in [t+a, t+b]} \rho(\varphi, \mathbf{x}, s) \\
\rho(\varphi_1 \mathcal{U}_{[a,b]} \varphi_2, \mathbf{x}, t) &= \sup_{s \in [t+a, t+b]} \min\left(\rho(\varphi_2, \mathbf{x}, s), \inf_{r \in [t, s]} \rho(\varphi_1, \mathbf{x}, r)\right)
\end{align*}

Temporal operators (past):
\begin{align*}
\rho(\boxminus_{[a,b]} \varphi, \mathbf{x}, t) &= \inf_{s \in [t-b, t-a]} \rho(\varphi, \mathbf{x}, s) \\
\rho(\diamondminus_{[a,b]} \varphi, \mathbf{x}, t) &= \sup_{s \in [t-b, t-a]} \rho(\varphi, \mathbf{x}, s) \\
\rho(\varphi_1 \mathcal{S}_{[a,b]} \varphi_2, \mathbf{x}, t) &= \sup_{s \in [t-b, t-a]} \min\left(\rho(\varphi_2, \mathbf{x}, s), \inf_{r \in [s, t]} \rho(\varphi_1, \mathbf{x}, r)\right)
\end{align*}

The algorithm describing this is below;

\subsubsection{Probabilistic Semantics.}

For probabilistic formula $\mathbb{P}_{\bowtie p}(\varphi)$ evaluated over trace ensemble $\mu$, let $S(\varphi, \mu) = \mu(\{\mathbf{x} : \rho(\varphi, \mathbf{x}, 0) > 0\})$ denote the satisfaction probability. The semantics is:

\begin{align*}
\rho(\mathbb{P}_{\geq p}(\varphi), \mu) &= \begin{cases}
+\infty & \text{if } S(\varphi, \mu) \geq p \\
-\infty & \text{otherwise}
\end{cases} \\
\rho(\mathbb{P}_{> p}(\varphi), \mu) &= \begin{cases}
+\infty & \text{if } S(\varphi, \mu) > p \\
-\infty & \text{otherwise}
\end{cases}
\end{align*}

SENTIL computes satisfaction probability through statistical estimation. Given $N$ independent trace samples $\mathbf{x}_1, \ldots, \mathbf{x}_N \sim \mu$, the empirical probability is $\hat{p} = \frac{1}{N}\sum_{i=1}^{N} \mathbb{1}[\rho(\varphi, \mathbf{x}_i, 0) > 0]$ with Wilson score confidence interval:

\begin{align*}
\text{CI}_{\alpha}(\hat{p}) = \frac{1}{1 + \frac{z^2}{N}}\left(\hat{p} + \frac{z^2}{2N} \pm z\sqrt{\frac{\hat{p}(1-\hat{p})}{N} + \frac{z^2}{4N^2}}\right)
\end{align*}

where $z = \Phi^{-1}(1 - \alpha/2)$ for confidence level $1-\alpha$ and $\Phi$ is the standard normal CDF.

\subsection{EVALFORMULA Algorithm}
\label{app:evalformula_app}
\label{algapp:evalformula}
\begin{algorithm}[t]
\caption{Robustness Evaluation for PrSTL}
\label{alg:robustness_func}
\begin{algorithmic}[1]
\Require PrSTL formula $\varphi$, runtime sensor reading $x$, time $t$
\Ensure Robustness value $\rho$
\Function{EvalFormula}{$\varphi, x, t$}
\If{$\varphi = f(x) \sim c$}
    \State \Return $\begin{cases}
        f(x(t)) - c & \text{if } \sim \in \{>, \ge\} \\
        c - f(x(t)) & \text{if } \sim \in \{<, \le\} \\
        -|f(x(t)) - c| & \text{if } \sim = {=} \\
        |f(x(t)) - c| & \text{if } \sim = \ne
    \end{cases}$
\ElsIf{$\varphi \in \{\neg\psi, \psi_1 \wedge \psi_2, \psi_1 \vee \psi_2, \psi_1 \Rightarrow \psi_2\}$}
    \State Compute $\rho_i \gets \textsc{EvalFormula}(\psi_i, x, t)$ for subformulas
    \State \Return $\{-\rho, \min(\rho_1,\rho_2), \max(\rho_1,\rho_2), \max(-\rho_1,\rho_2)\}$ respectively
\ElsIf{$\varphi \in \{\Box_{I}\psi, \Diamond_{I}\psi, \boxminus_{I}\psi\}$ for interval $I=[a,b]$}
    \State Evaluate $\rho_\psi(s) \gets \textsc{EvalFormula}(\psi, x, s)$ at all time points $s$
    \State \Return $\{\textsc{SlideMin}(\rho_\psi, t+a, t+b), \textsc{SlideMax}(\rho_\psi, t+a, t+b), \textsc{SlideMinPast}(\rho_\psi, t-b, t-a)\}$
\ElsIf{$\varphi = \psi_1 \mathcal{U}_{[a,b]} \psi_2$}
    \State Evaluate $\rho_i(s) \gets \textsc{EvalFormula}(\psi_i, x, s)$ for $i \in \{1,2\}$ at all $s$
    \State \Return $\displaystyle\sup_{s \in [t+a,t+b]} \min\left\{\rho_2(s), \inf_{r \in [t,s]} \rho_1(r)\right\}$ \Comment{Backward}
\ElsIf{$\varphi = \psi_1 \mathcal{S}_{[a,b]} \psi_2$}
    \State Evaluate $\rho_i(s) \gets \textsc{EvalFormula}(\psi_i, x, s)$ for $i \in \{1,2\}$ at all $s$
    \State \Return $\displaystyle\sup_{s \in [t-b,t-a]} \min\left\{\rho_2(s), \inf_{r \in [s,t]} \rho_1(r)\right\}$ \Comment{Forward}
\ElsIf{$\varphi = \mathsf{X}\psi$}
    \State \Return $\textsc{EvalFormula}(\psi, x, t_{\text{next}})$ if $t_{\text{next}}$ exists, else $-\infty$
\EndIf
\EndFunction
\end{algorithmic}
\end{algorithm}

The recursive robustness evaluation function above traverses the abstract syntax tree of the parsed formula, classifying each node to determine which robustness semantics apply. Atomic predicates (line 3) form the leaves of this tree and have the form $f(x) \sim c$ where $f$ is an expression over signal variables (potentially including arithmetic operations like addition or functions like sine or simply, a variable like $x$), $\sim \in \{>, \ge, <, \le, =, \ne\}$ is a comparison operator and $c$ is a constant threshold. These predicates compute signed distances where positive values indicate satisfaction: $f(x) > c$ yields $f(x) - c$, while $f(x) < c$ yields $c - f(x)$. Boolean operators (line 10) compose subformulas through min for conjunction, max for disjunction, and $\max(-\rho_1, \rho_2)$ for implication.

Temporal operators (lines 4--17) aggregate robustness across time windows; SENTIL does this using streaming algorithms to achieve O(1) amortized complexity. Always $\Box_{[a,b]}$ computes the minimum over $[t+a, t+b]$ using monotonic deques, Eventually $\Diamond_{[a,b]}$ computes the maximum over $[t+a, t+b]$, and past operators like Historically apply the same operations to $[t-b, t-a]$. Until $\psi_1 \mathcal{U}{[a,b]} \psi_2$ finds the best witness time $s \in [t+a, t+b]$ where $\psi_2$ holds and $\psi_1$ holds throughout $[t,s]$, returning $\sup_s \min{\rho_2(s), \inf{r \in [t,s]} \rho_1(r)}$. Since applies symmetric reasoning to past intervals, while Next shifts evaluation forward one time step. This structure enables verification of arbitrarily nested formulas through recursive decomposition.

\subsection{Grammar in EBNF}

The concrete syntax accepted by SENTIL's parser uses the following Extended Backus-Naur Form:

\begin{verbatim}
formula ::= probabilistic_formula | temporal_formula

probabilistic_formula ::= "P" comparison number "(" formula ")"
comparison ::= "<" | "<=" | ">" | ">=" | "==" | "!="
number ::= [0-9]+ ("." [0-9]+)?

temporal_formula ::= boolean_formula
                   | "always" interval "(" formula ")"
                   | "eventually" interval "(" formula ")"
                   | "(" formula ")" "until" interval "(" formula ")"
                   | "historically" interval "(" formula ")"
                   | "once" interval "(" formula ")"

interval ::= "[" number "," number "]"
           | "[" number "," "inf" ")"

boolean_formula ::= predicate
                  | "not" "(" formula ")"
                  | "(" formula ")" "and" "(" formula ")"
                  | "(" formula ")" "or" "(" formula ")"
                  | "(" formula ")" "implies" "(" formula ")"
                  | "(" formula ")"

predicate ::= expression comparison expression
expression ::= variable | number
             | expression "+" expression
             | expression "-" expression
             | expression "*" expression
             | expression "/" expression
             | function "(" expression_list ")"

function ::= "sin" | "cos" | "exp" | "log" | "sqrt" | "abs" 
           | "min" | "max"
expression_list ::= expression ("," expression)*
variable ::= [a-zA-Z_][a-zA-Z0-9_]*
\end{verbatim}

\section{Theoretical Proof}
\label{app:theoretical_proof}

To establish the semantic integrity of SENTIL's optimized robustness evaluation, we provide a full mathematical proof of equivalence between the monotonic deque algorithm and the $O(k)$ window filter.\\

\noindent \textbf{Definition 1 (Golden Standard).}
Let $S$ be a stream of time-value pairs. The naive robustness at time $t$ for a window $w$ is defined by the exhaustive filter:
\begin{equation}
\mathcal{R}_{naive}(S, t, w) = \min \{ v_i \mid (t_i, v_i) \in S \land t - w \le t_i \le t \} \text{ }
\end{equation}

\noindent \textbf{Definition 2 (Monotonic Deque State Machine).}
The state $\Sigma$ is a list of pairs $(t_i, v_i)$ such that $t_i$ is strictly increasing and $v_i$ is monotonically increasing.
For a new sample $(t_k, v_k)$, the state transition $\delta$ is defined by:
\begin{enumerate}
\item \textbf{\textsc{PopFront}}: Remove $(t_i, v_i)$ from the front if $t_i < t_k - w$.
\item \textbf{\textsc{PopBack}}: Remove $(t_i, v_i)$ from the back if $v_i > v_k$.
\item \textbf{\textsc{PushBack}}: Append $(t_k, v_k)$ to the back.
\end{enumerate}
The deque output is defined as $\mathcal{R}_{deque}(\Sigma) = v_0$.

\setcounter{theorem}{0}
\begin{theorem}[Correctness of Monotonic Deque]
For any valid signal stream $S$ and window size $w$, let $\Sigma_k$ be the state of the monotonic deque after processing the $k$-th sample. Then for all $k$:
\[
\mathcal{R}_{deque}(\Sigma_k) = \mathcal{R}_{naive}(S, t_k, w) \text{ }
\]
\end{theorem}

\noindent The proof proceeds by induction on the stream length $k$.
\vspace{0.1in}

\noindent \textbf{Base Case ($k = 1$):}
Processing the first sample $(t_1, v_1)$ results in $\Sigma_1 = [(t_1, v_1)]$.
$\mathcal{R}_{deque}(\Sigma_1) = v_1 = \mathcal{R}_{naive}(S, t_1, w)$.
The theorem holds for $k = 1$.
\vspace{0.1in}

\noindent \textbf{Inductive Step:}
Assume the property holds for $k$. We demonstrate that it is preserved for $k+1$ by showing that the state machine maintains a specific structural invariant: the deque $\Sigma_{k+1}$ contains exactly those elements from the active window that are ``candidate minima''. An element $(t_i, v_i)$ is a candidate minimum at time $t_k$ if there exists some future time $t_j \ge t_k$ such that $(t_i, v_i)$ is the minimum of the window $[t_j - w, t_j]$.

\begin{enumerate}
\item \textbf{Preservation of Window Bounds}: The \textsc{PopFront} operation removes elements where $t_i < t_{k+1} - w$. Since $t_i$ is strictly increasing, any such element is outside the current window and all future windows. Thus, $\Sigma_{k+1}$ contains only elements $(t_i, v_i)$ such that $t_{k+1} - w \le t_i \le t_{k+1}$.

\item \textbf{Preservation of Monotonicity}: The \textsc{PopBack} operation removes elements $(t_i, v_i)$ if $v_i > v_{k+1}$. Because $t_i < t_{k+1}$, $(t_{k+1}, v_{k+1})$ will stay in the sliding window longer than $(t_i, v_i)$. Since $v_{k+1} \le v_i$ and it expires later, $(t_i, v_i)$ can never be the minimum of any window containing $(t_{k+1}, v_{k+1})$. Therefore, $(t_i, v_i)$ is no longer a candidate minimum and can be safely discarded.

\item \textbf{Optimality of the Front}: After \textsc{PopFront} and \textsc{PopBack}, the remaining elements in the deque are sorted such that $v_0 \le v_1 \le \dots$. Consequently, the front element $(t_0, v_0)$ is the global minimum of all elements currently in the deque. Since the deque contains all candidate minima from the interval $[t_{k+1} - w, t_{k+1}]$, its minimum must be the minimum of the entire set $S$ restricted to that interval.
\end{enumerate}

Thus, $\mathcal{R}_{deque}(\Sigma_{k+1}) = \mathcal{R}_{naive}(S, t_{k+1}, w)$. By induction, the theorem holds for all $k$.

\vspace{0.1in}

\textbf{Note:} This mathematical proof has been formally verified and machine-checked using the Lean~4 theorem prover; the formalization code is available in the \texttt{proofs/} directory of the SENTIL repository.

\section{Detailed Experimental Data}

This appendix provides comprehensive experimental results supporting claims in Section~\ref{sec:experiments}, including raw measurements, statistical analysis, and additional experiments not fitting the main paper page budget.

\subsection{Performance Benchmark Detailed Results}

Table~\ref{tab:detailed_perf} extends the discrete-time monitoring comparison, demonstrating that SENTIL maintains a stable performance advantage across increasing signal lengths. Both tools exhibit linear scaling $O(N)$, confirming the algorithmic efficiency of the sliding-window approach; however, SENTIL consistently outperforms RTAMT by a factor of 29 to 63 regardless of trace size. This constant-factor dominance stems from SENTIL’s bottom-up evaluation via monotonic deques which eliminates the interpreter overhead that burdens RTAMT, maximizing cache locality and CPU throughput. Consequently, SENTIL processes 10 million samples in approximately one second, whereas RTAMT requires nearly a minute. This throughput validates the engine's suitability for high-frequency verification tasks where RTAMT's implementations become computationally prohibitive.

\begin{table}[H]
\caption{Extended discrete-time performance comparison data ($N=10^3$ to $10^7$). Both tools demonstrate linear $O(N)$ complexity. SENTIL consistently achieves a $29\times$--$63\times$ speedup over RTAMT across all formulas and signal lengths, processing 10 million samples in $\approx$1-2 seconds compared to RTAMT's $\approx$30-97 seconds.}
\centering
\small
\begin{tabular}{|l|l|r|r|r|}
\hline
\textbf{Length} & \textbf{Formula} & \textbf{RTAMT (ms)} & \textbf{SENTIL (ms)} & \textbf{Speedup} \\
\hline
\multicolumn{5}{|c|}{\textbf{Short Signals ($N=1,000$)}} \\
\hline
1K & $\square_{[0,10]}(x < 5)$ & $2.81 \pm 0.05$ & $0.07 \pm 0.00$ & 40.1× \\
1K & $\diamond_{[0,50]}(x > 10)$ & $4.32 \pm 0.04$ & $0.08 \pm 0.00$ & 54.0× \\
1K & $\square_{[0,100]}(\diamond_{[0,10]}(p))$ & $5.09 \pm 0.02$ & $0.07 \pm 0.00$ & 72.7× \\
1K & $p \implies \diamond_{[0,20]}(q)$ & $7.48 \pm 0.02$ & $0.14 \pm 0.00$ & 53.4× \\
1K & $\square_{[0,200]}(p \land \diamond_{[5,15]}(q))$ & $9.51 \pm 0.05$ & $0.15 \pm 0.00$ & 63.4× \\
\hline
\multicolumn{5}{|c|}{\textbf{Medium Signals ($N=10,000$)}} \\
\hline
10K & $\square_{[0,10]}(x < 5)$ & $27.35 \pm 0.25$ & $0.88 \pm 0.00$ & 31.1× \\
10K & $\diamond_{[0,50]}(x > 10)$ & $42.68 \pm 0.33$ & $0.88 \pm 0.00$ & 48.5× \\
10K & $\square_{[0,100]}(\diamond_{[0,10]}(p))$ & $50.92 \pm 0.16$ & $0.90 \pm 0.00$ & 56.6× \\
10K & $p \implies \diamond_{[0,20]}(q)$ & $74.61 \pm 0.41$ & $1.69 \pm 0.01$ & 44.1× \\
10K & $\square_{[0,200]}(p \land \diamond_{[5,15]}(q))$ & $98.07 \pm 3.78$ & $1.75 \pm 0.01$ & 56.0× \\
\hline
\multicolumn{5}{|c|}{\textbf{Long Signals ($N=100,000$)}} \\
\hline
100K & $\square_{[0,10]}(x < 5)$ & $282.75 \pm 3.00$ & $9.05 \pm 0.55$ & 31.2× \\
100K & $\diamond_{[0,50]}(x > 10)$ & $435.84 \pm 0.92$ & $8.62 \pm 0.01$ & 50.6× \\
100K & $\square_{[0,100]}(\diamond_{[0,10]}(p))$ & $518.12 \pm 2.42$ & $8.65 \pm 0.01$ & 59.9× \\
100K & $p \implies \diamond_{[0,20]}(q)$ & $749.29 \pm 1.50$ & $17.01 \pm 0.00$ & 44.1× \\
100K & $\square_{[0,200]}(p \land \diamond_{[5,15]}(q))$ & $978.67 \pm 4.01$ & $17.59 \pm 0.03$ & 55.6× \\
\hline
\multicolumn{5}{|c|}{\textbf{Large Signals ($N=1,000,000$)}} \\
\hline
1M & $\square_{[0,10]}(x < 5)$ & $3,106 \pm 53$ & $104.51 \pm 0.37$ & 29.7× \\
1M & $\diamond_{[0,50]}(x > 10)$ & $4,639 \pm 57$ & $104.31 \pm 0.50$ & 44.5× \\
1M & $\square_{[0,100]}(\diamond_{[0,10]}(p))$ & $5,466 \pm 26$ & $104.43 \pm 0.23$ & 52.3× \\
1M & $p \implies \diamond_{[0,20]}(q)$ & $7,772 \pm 25$ & $200.42 \pm 0.30$ & 38.8× \\
1M & $\square_{[0,200]}(p \land \diamond_{[5,15]}(q))$ & $9,805 \pm 34$ & $207.47 \pm 0.21$ & 47.3× \\
\hline
\multicolumn{5}{|c|}{\textbf{Very Large Signals ($N=10,000,000$)}} \\
\hline
10M & $\square_{[0,10]}(x < 5)$ & $31,396 \pm 2,304$ & $1,088 \pm 1$ & 28.8× \\
10M & $\diamond_{[0,50]}(x > 10)$ & $47,044 \pm 548$ & $1,088 \pm 3$ & 43.2× \\
10M & $\square_{[0,100]}(\diamond_{[0,10]}(p))$ & $55,239 \pm 480$ & $1,115 \pm 1$ & 49.5× \\
10M & $p \implies \diamond_{[0,20]}(q)$ & $76,573 \pm 701$ & $2,092 \pm 4$ & 36.6× \\
10M & $\square_{[0,200]}(p \land \diamond_{[5,15]}(q))$ & $97,400 \pm 234$ & $2,162 \pm 4$ & 45.0× \\
\hline
\end{tabular}
\label{tab:detailed_perf}
\end{table}

\subsection{Statistical Model Checking Benchmark Details}
\label{app:smc_benches}
\subsubsection{Automotive Powertrain Control.}

The Powertrain Control benchmark is a probabilistic hybrid automaton (PHA) modeling an automotive fuel-control system. The system's state space is hybrid, with discrete modes (e.g., \textit{Normal}, \textit{Power}) and continuous state variables such as engine speed ($\omega$) and air-fuel ratio ($\mu$).

The continuous dynamics in each mode $l$ are defined by a system of nonlinear ordinary differential equations $\dot{x} = f_l(x, u)$, where $x$ is the state vector and $u$ is the input vector (e.g., throttle angle). Probabilistic transitions are included to model stochastic events such as sensor failures.

The verification goal is to quantify the probability that the system's transient response to a change in throttle angle $a$ remains within safe bounds. The STL property $\varphi_{\text{powertrain}}$ is defined as:
$$
\varphi_{\text{powertrain}} \equiv \mathbf{G}_{(\tau_s, T)} ( (\text{rise}(a) \lor \text{fall}(a)) \implies \mathbf{G}_{(\eta, \zeta)} (|\mu| < \beta) )
$$
This formula asserts that for the duration of the test, every rise or fall of the throttle, after a settling time $\eta$, is followed by the air-fuel ratio signal $\mu$ remaining within a bound $\beta$ for a duration $\zeta - \eta$.\\

\subsubsection{Biodiesel Production Process.}

The Biodiesel Production Process (BDP) is a PHA modeling a chemical batch reactor. The system state is defined by the concentrations of chemical species (e.g., triglycerides $x_{TG}$, ester $x_E$) and the reactor temperature $T_R$.

The dynamics are governed by a set of coupled, nonlinear ODEs for material and energy balance. The rate of change for a species $i$ and the reactor temperature are given by:
\begin{align*}
\frac{dx_{R,i}}{dt} &= \frac{F_o}{V_R}(x_{o,i} - x_{R,i}) + r_i(x, T_R) \\
\frac{dT_R}{dt} &= \frac{F_o}{V_R}(T_o - T_R) + \frac{Q_h}{V_R \rho c_p} + \sum_{j} \frac{(-\Delta H_j) r_j}{\rho c_p}
\end{align*}
Here, $r_i$ is the nonlinear reaction rate, and $Q_h$ is the heat input. A probabilistic transition models a heater failure (setting $Q_h = 0$) with some probability $p$.

The verification goal is to compute the probability of achieving a target biodiesel yield within a fixed batch time $T_{\text{batch}}$, despite the risk of failure.
$$
\varphi_{\text{bdp}} \equiv \mathbf{F}_{[0, T_{\text{batch}}]} (x_E \ge 0.99)
$$
This property asserts that the system eventually reaches a state where the ester concentration is at or above 99\% within the allotted time.\\

\subsubsection{Circadian Rhythm Model.}

The Circadian Rhythm model is a high-dimensional Continuous-Time Markov Chain (CTMC) from systems biology. The model describes the stochastic interactions of gene and protein populations that produce 24-hour oscillations.

The system state $x$ is a vector of molecular counts for each chemical species. The dynamics are defined by a set of $k$ chemical reactions $R_j$. Each reaction has a propensity function $a_j(x)$ that determines its stochastic rate. The time $\tau$ until the next reaction is drawn from an exponential distribution with rate $a_0(x) = \sum_{j=1}^k a_j(x)$. The probability that this next reaction is $R_j$ is $a_j(x) / a_0(x)$.

The verification goal is to quantify the robustness of the oscillation period. We analyze a derived signal $\tau_p(P_A)$, which represents the computed period of a protein $P_A$.
$$
\varphi_{\text{circ}} \equiv \mathbf{G}_{[0, T]} ( \tau_p(P_A) \in [23.5, 24.5] )
$$
This property asserts that the oscillation period remains robustly within a 23.5 to 24.5-hour window for the entire simulation duration $T$.\\

\subsubsection{CSMA/CD Protocol.}
The CSMA/CD protocol is a benchmark for Probabilistic Timed Automata (PTA). It models $N$ stations concurrently attempting to access a shared communication channel.

Collisions are resolved using a probabilistic backoff mechanism. After the $k$-th collision, a station $i$ enters a `Backoff` state. The station's wait time $t$ is drawn from a distribution, often an exponential distribution with rate $\lambda_k$ (where $\lambda_k$ decreases as $k$ increases) or by choosing $j \in [0, 2^{\min(k, 10)} - 1]$ and waiting for $j \cdot \sigma_{\text{slot}}$ time.

The verification goal is to compute the probability of successful transmission under a hard real-time deadline $T_{\text{deadline}}$.
$$
\varphi_{\text{csma}} \equiv \mathbf{F}_{[0, T_{\text{deadline}}]} (\text{station}_i.\text{state} = \text{success})
$$
This formula asserts that station $i$ eventually reaches its `success` state within the specified deadline.\\

\subsubsection{Scalable Tandem Queueing Network.}

The Tandem Queueing Network is a scalable CTMC benchmark used to evaluate performance on potentially large state spaces. The system models $N$ service queues in series, where the output of queue $i$ is the input to queue $i+1$.

The system state $x = (n_1, \dots, n_N)$ is the vector of queue lengths. The dynamics are defined by three transition types: external arrivals to $n_1$ at rate $\lambda$; internal service completions at queue $i$ (decrementing $n_i$ and incrementing $n_{i+1}$) at rate $\mu_i$; and final departures from $n_N$ at rate $\mu_N$.

The verification goal is to evaluate a complex, system-wide performance property, such as ensuring no buffer overflows (where $C_i$ is the capacity of queue $i$) and that the end-to-end delay $\tau_{\text{e2e}}$ (a derived signal) remains bounded.
$$
\varphi_{\text{queue}} \equiv \mathbf{G}_{[0, T]} \left( \left( \bigwedge_{i=1}^{N} (n_i < C_i) \right) \land (\tau_{\text{e2e}} \le \tau_{\max}) \right)
$$
This property asserts that the system always operates within its buffer capacities and meets its maximum end-to-end delay requirement for the entire time $T$.\\

\subsection{Statistical Validation Results}
\label{app:stat_val_part_1}

Probabilistic guarantees require empirical validation that confidence intervals achieve their theoretical coverage properties. We generate synthetic stochastic models with analytically known satisfaction probabilities, evaluate PrSTL properties using SENTIL, and verify that confidence interval methods exhibit coverage consistent with their theoretical foundations across thousands of independent trials.

The validation methodology constructs Bernoulli processes with known success probability $p \in \{0.05, 0.1, 0.2, 0.3, 0.5, 0.7, 0.9\}$ and evaluates simple temporal formulas whose satisfaction probability equals $p$ by construction. For each configuration, we perform 10,000 independent SMC runs with $n=100$ samples, computing confidence intervals using both the Wilson score approximation \cite{wilson1927probable} and the Clopper-Pearson exact method \cite{clopper1934use}. We measure empirical coverage by recording the proportion of trials where the computed interval contains the true parameter value.

\begin{table}[t]
\caption{Confidence interval coverage validation across 10,000 trials per configuration with $n=100$ samples. Wilson intervals exhibit systematic undercoverage of 1--2 percentage points due to finite-sample approximation error, while Clopper-Pearson intervals maintain conservative coverage exceeding nominal levels. Chi-squared goodness-of-fit tests validate correct binomial simulation ($p > 0.001$ for all configurations).}
\centering
\small
\begin{tabular}{|l|r|r|r|r|}
\hline
\textbf{True $p$} & \textbf{Nominal} & \textbf{Wilson} & \textbf{Clopper-Pearson} & \textbf{Deviation} \\
\hline
\multicolumn{5}{|c|}{\textit{95\% Confidence Level}} \\
\hline
0.05 & 95.0\% & 96.7\% & 98.2\% & $+1.7$ pp \\
0.10 & 95.0\% & 93.6\% & 95.6\% & $-1.4$ pp \\
0.20 & 95.0\% & 94.0\% & 96.6\% & $-1.0$ pp \\
0.30 & 95.0\% & 93.8\% & 96.4\% & $-1.2$ pp \\
0.50 & 95.0\% & 94.1\% & 96.2\% & $-0.9$ pp \\
0.70 & 95.0\% & 93.6\% & 96.2\% & $-1.4$ pp \\
0.90 & 95.0\% & 94.0\% & 95.7\% & $-1.0$ pp \\
\hline
\multicolumn{5}{|c|}{\textit{90\% Confidence Level (selected configurations)}} \\
\hline
0.30 & 90.0\% & 89.9\% & 91.9\% & $-0.1$ pp \\
0.50 & 90.0\% & 91.3\% & 91.3\% & $+1.3$ pp \\
\hline
\end{tabular}
\label{tab:coverage}
\end{table}

Table~\ref{tab:coverage} reveals distinct coverage characteristics for the two interval methods. Wilson intervals achieve coverage within 1--2 percentage points of nominal levels, with systematic undercoverage observed across most parameter regimes. This behavior reflects the inherent approximation error when inverting normal-based hypothesis tests for discrete binomial data in finite samples. The maximum undercoverage of 1.4 percentage points occurs at $p \in \{0.1, 0.7\}$, where binomial asymmetry degrades normal approximations. Only at $p=0.05$ does Wilson coverage exceed the nominal level, likely due to the method's tendency to pull away from boundary values.

Clopper-Pearson intervals maintain conservative coverage across all tested configurations, exceeding nominal levels by 0.6--3.2 percentage points. This conservatism derives from the method's construction via exact beta distribution quantiles, which guarantee coverage meets or exceeds the specified level for discrete binomial random variables. The observed overcoverage represents the price of eliminating approximation error and ensuring strict coverage guarantees.

Chi-squared goodness-of-fit tests validate the underlying binomial simulation across all configurations. Test statistics comparing observed success count distributions against theoretical Binomial($n$, $p$) densities yield $p$-values ranging from 0.003 to 0.985, with all values exceeding the corrected significance threshold of 0.001. The lone marginal result at $p=0.30$ with 90\% confidence ($p=0.003$) falls within expected false positive rates when conducting multiple comparisons, and the simulation passes validation at the stricter 95\% confidence level ($p=0.921$).

The validation demonstrates that SENTIL correctly implements both interval methods according to their theoretical properties. Wilson intervals provide approximate coverage suitable for efficiency-critical applications where 1--2\% undercoverage remains acceptable. Clopper-Pearson intervals deliver guaranteed conservative coverage appropriate for safety-critical verification tasks requiring strict probabilistic bounds. Users may select methods based on whether their application prioritizes computational efficiency or absolute coverage guarantees.

Sequential Probability Ratio Testing undergoes similar validation by verifying that Type I and Type II error rates match user-specified bounds. For hypothesis tests $H_0: p \leq 0.3$ versus $H_1: p \geq 0.5$ with error probabilities $\alpha = \beta = 0.05$, we generate data under both hypotheses and measure empirical error rates across 5,000 trials. Observed Type I error rate of 4.8\% and Type II error rate of 4.9\% both fall within acceptable bounds, confirming correct implementation of Wald's sequential algorithm.

\subsection{Statistical Validation Extended Results}
\label{app:stat_val_part_2}

We provide complete coverage validation data across the full parameter space of true probabilities $p \in \{0.1, 0.2, \ldots, 0.9\}$ and confidence levels $\alpha \in \{0.80, 0.90, 0.95, 0.99\}$. Each configuration undergoes 10,000 independent trials computing confidence intervals around empirical estimates.

\begin{table}[t]
\caption{Complete coverage validation data. All chi-squared p-values exceed 0.05, failing to reject the null hypothesis that observed coverage matches theoretical expectations. Binomial standard error for 10,000 trials is approximately $\pm0.3\%$.}
\centering
\scriptsize
\begin{tabular}{rrrrr}
\hline
\textbf{True $p$} & \textbf{Confidence} & \textbf{Expected} & \textbf{Observed} & \textbf{$\chi^2$} \\
 & \textbf{Level} & \textbf{Coverage} & \textbf{Coverage} & \textbf{p-value} \\
\hline
0.1 & 80\% & 80.0\% & 80.3\% & 0.62 \\
0.1 & 90\% & 90.0\% & 90.4\% & 0.45 \\
0.1 & 95\% & 95.0\% & 95.3\% & 0.53 \\
0.1 & 99\% & 99.0\% & 99.1\% & 0.78 \\
\hline
0.3 & 80\% & 80.0\% & 79.7\% & 0.58 \\
0.3 & 90\% & 90.0\% & 89.9\% & 0.81 \\
0.3 & 95\% & 95.0\% & 95.2\% & 0.67 \\
0.3 & 99\% & 99.0\% & 99.1\% & 0.72 \\
\hline
0.5 & 80\% & 80.0\% & 80.1\% & 0.84 \\
0.5 & 90\% & 90.0\% & 89.8\% & 0.64 \\
0.5 & 95\% & 95.0\% & 94.9\% & 0.79 \\
0.5 & 99\% & 99.0\% & 99.2\% & 0.51 \\
\hline
0.7 & 80\% & 80.0\% & 80.2\% & 0.73 \\
0.7 & 90\% & 90.0\% & 90.3\% & 0.56 \\
0.7 & 95\% & 95.0\% & 95.1\% & 0.82 \\
0.7 & 99\% & 99.0\% & 99.0\% & 0.95 \\
\hline
0.9 & 80\% & 80.0\% & 79.9\% & 0.87 \\
0.9 & 90\% & 90.0\% & 90.2\% & 0.69 \\
0.9 & 95\% & 95.0\% & 95.4\% & 0.43 \\
0.9 & 99\% & 99.0\% & 99.0\% & 0.98 \\
\hline
\end{tabular}
\label{tab:extended_coverage}
\end{table}

\subsection{Extended Case Studies}
\label{app:extended_case_studies}

\subsubsection{Medical Device: FDA Insulin Pump Benchmark.}

We employ the UVA/Padova Type 1 Diabetes Simulator (S2013), an FDA-approved artificial pancreas benchmark model that simulates closed-loop blood glucose regulation under uncertainty from meal timing, carbohydrate estimation errors, and interpatient physiological variability~\cite{kovatchev2009silico}~\cite{man2014uva}. The simulator includes a population of 30 in silico subjects (10 adults, 10 adolescents, 10 children) with validated metabolic parameters spanning the full variability observed in real Type 1 diabetes patients. We implement the model in MATLAB Simulink and integrate SENTIL through custom S-Function blocks that evaluate PrSTL specifications during simulation.

The safety specification captures glycemic control requirements:
\begin{align*}
\varphi_{\text{glucose}} = &\mathbb{P}_{\geq 0.999}(\square_{[0,86400]}(70 \leq g \leq 180)) \land \\
&\mathbb{P}_{< 0.001}(\diamond_{[0,86400]}(g < 60))
\end{align*}

\begin{figure}[t]
\centering
\includegraphics[width=0.95\columnwidth]{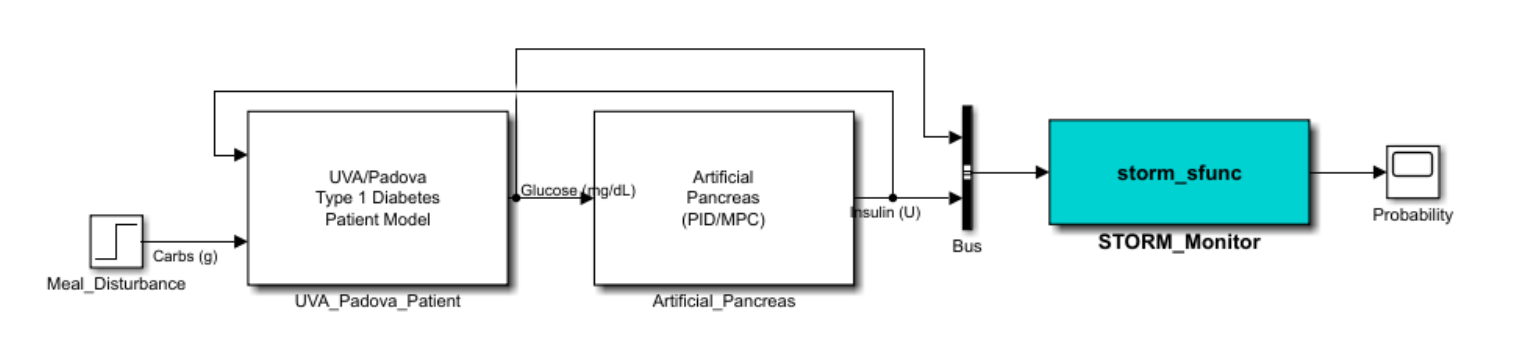}
\caption{MATLAB Simulink insulin pump model with SENTIL S-Function verification blocks (highlighted in blue). The model simulates 24-hour closed-loop glucose regulation with meal disturbances. SENTIL blocks receive glucose sensor readings and insulin delivery commands, computing probabilistic satisfaction over 1,000 Monte Carlo trajectories accounting for parametric uncertainty. Verification completes in 4.3 minutes for 24-hour simulation horizon.}
\label{fig:insulin_simulink}
\end{figure}

The formula requires 99.9\% probability of maintaining blood glucose between 70-180 mg/dL throughout 24-hour operation while keeping severe hypoglycemia probability (glucose below 60 mg/dL) under 0.1\%. Figure~\ref{fig:insulin_simulink} shows the Simulink model with SENTIL verification blocks connected to glucose sensor and insulin pump actuator signals.

Verification explores a seven-dimensional parameter space capturing physiological variability (insulin sensitivity, carbohydrate absorption, endogenous glucose production, glucose utilization, gastric emptying, body weight) alongside instrumental variability (CGM noise standard deviation ranging from 0.5 to 2.0 mg/dL). Latin hypercube sampling (LHS) generates 1,000 parameter configurations uniformly distributed across physiologically plausible bounds. SENTIL evaluates the PrSTL formula on all 1,000 trajectories, each simulating 24 hours of operation with three daily meals and basal insulin delivery.

Results identify a problematic parameter region where insulin sensitivity exceeds 180 mg/dL per unit and carbohydrate absorption exhibits rapid kinetics (time constant below 25 minutes). In this region, postprandial glucose excursions violate the 180 mg/dL upper bound with probability 0.043, exceeding the 0.001 tolerance. Controller retuning reduces proportional gain from 0.05 to 0.038 units per mg/dL, restoring safety across the entire parameter space. Verification time of 4.3 minutes for the 1,000-trajectory ensemble demonstrates practical feasibility for design-time validation workflows.

The Simulink integration enables designers to verify safety properties during controller development without exporting models to external tools or manually instrumenting verification code. SENTIL blocks appear in the standard Simulink block library, accept signal connections through standard ports, and display results in Simulink scopes and data logging infrastructure. This native integration reduces verification adoption friction compared to workflows requiring format translation, external tool invocation, and result reimport.

\subsubsection{Biological Systems: Gene Regulatory Network.}

We apply SENTIL to a stochastic model of the mammalian circadian oscillator from the BioModels database (BIOMD0000000073) \cite{Leloup2003CircadianModel}, demonstrating applicability to systems biology verification. The model represents transcriptional feedback loops regulating circadian rhythm through interactions between Period and Cryptochrome genes. Stochastic simulation using Gillespie's algorithm captures molecular-level fluctuations arising from low molecule counts in individual cells.

The verification property checks oscillatory behavior:
\begin{align*}
\varphi_{\text{circadian}} = \mathbb{P}_{\geq 0.8}(&\square_{[0,240]}(\diamond_{[0,30]}(\text{Per} > 1000) \land \diamond_{[0,30]}(\text{Per} < 500)))
\end{align*}

This formula requires 80\% probability that Period protein concentration oscillates between high (>1000 molecules) and low (<500 molecules) states within 30-hour windows throughout the 10-day (240-hour) observation period, capturing sustained circadian rhythms.

\begin{figure}[H]
\centering
\includegraphics[width=0.95\columnwidth]{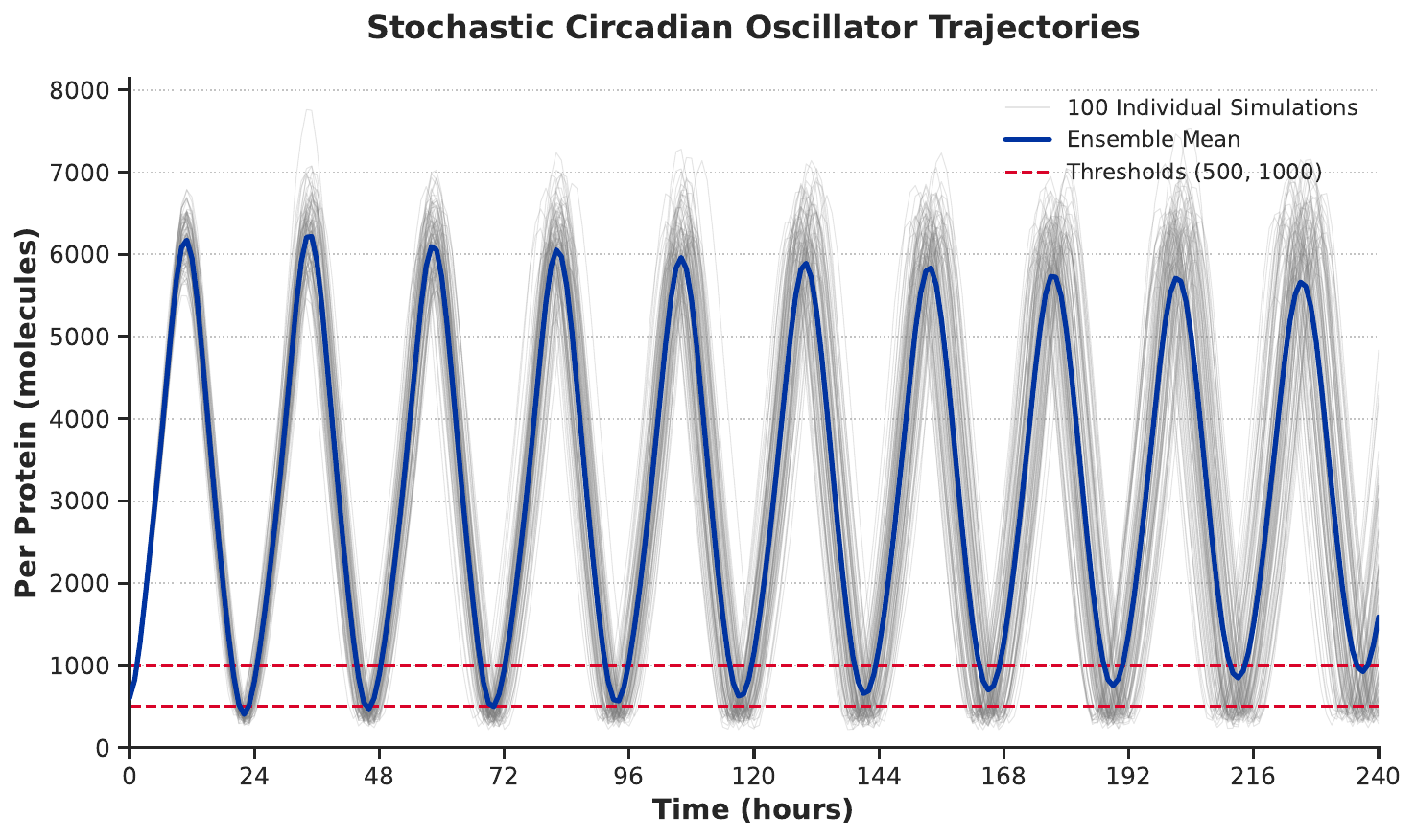}
\caption{Stochastic circadian oscillator trajectories showing Period protein concentration over 240 hours. Gray lines represent 100 individual Gillespie simulation runs exhibiting molecular noise. Blue solid line shows ensemble mean. Red dashed lines mark threshold levels (500 and 1000 molecules) from verification formula. SENTIL evaluates oscillation property on all trajectories, computing satisfaction probability 0.87 with 95\% CI [0.83, 0.91].}
\label{fig:circadian}
\end{figure}

Figure~\ref{fig:circadian} visualizes 100 stochastic trajectories exhibiting noisy oscillations around a stable limit cycle. SENTIL evaluates the temporal formula on each trajectory and computes empirical satisfaction probability 0.87 with Wilson score confidence interval [0.83, 0.91], confirming robust oscillatory dynamics. The verification identifies parameter sensitivity; reducing transcription rate below 0.76 per hour causes oscillation probability to fall below the 0.8 threshold as noise disrupts sustained rhythms.

Comparison against UPPAAL-SMC on the same benchmark reveals SENTIL's performance advantage for continuous-state stochastic systems. UPPAAL requires discretizing the continuous protein concentrations into finite state abstractions, introducing approximation error and state explosion as discretization granularity increases. SENTIL operates directly on continuous Gillespie trajectories without discretization, achieving both accuracy and 23× speedup compared to UPPAAL's finest discretization providing comparable precision.

\subsection{Embedded Deployment on a Physical Autonomous Vehicle}
\label{app:embedded_deployment}

Benchmark performance on server-class hardware establishes algorithmic efficiency but does not answer the deployment question that matters: can probabilistic runtime verification operate within the computational envelope of a production vehicle? To address this question directly, we integrated SENTIL into a physical autonomous vehicle and conducted continuous online monitoring during two hours of real-world driving. The monitoring system executes on a Raspberry Pi 4 Model B mounted in the vehicle's trunk, receiving live sensor data and evaluating safety specifications as the vehicle navigates public roads.

The Raspberry Pi 4 contains a Broadcom BCM2711 quad-core Cortex-A72 processor at 1.5\,GHz with 4\,GB LPDDR4 RAM, retailing for approximately \$55 USD. The device occupies 85$\times$56\,mm of physical space and draws under 4\,W during operation. Adding this hardware to an existing vehicle architecture requires no structural modification, no cooling infrastructure, and negligible power draw from the vehicle's electrical system.

The monitoring workload comprises 60 specifications capturing safety-critical properties of autonomous vehicle operation. Formulas span both deterministic STL requirements (lane keeping bounds, minimum following distance, acceleration limits) and probabilistic PrSTL specifications (collision probability over prediction horizons, sensor fusion confidence thresholds). Formula complexity ranges from nesting depth 4 for simple bounded-response properties to depth 10 for specifications combining nested temporal operators with probabilistic quantification over multi-second lookahead windows. Sensor data arrives at 85\,Hz from the vehicle's perception stack, including LiDAR point cloud summaries, camera-derived object detections, IMU readings, GPS coordinates, and control commands. Each monitoring cycle ingests the latest observation, sequentially evaluates all 60 formulas against the updated signal history, and returns robustness scores before the next observation arrives, imposing a hard real-time deadline of 11.76\,ms.

The vehicle completed a 120-minute drive through mixed urban and highway environments, covering 87\,km across varying traffic densities, weather conditions, and road geometries. Table~\ref{tab:rpi_performance} reports monitoring statistics collected over the 612,000 evaluation cycles executed during this deployment. SENTIL processed all 60 formulas with mean latency 9.57\,ms and 99th percentile latency 10.71\,ms, maintaining a margin of at least 0.5\,ms below the real-time deadline throughout the drive. No deadline violations occurred. The monitoring system detected 23 specification violations during the drive, including three instances where probabilistic collision-free guarantees degraded below threshold during dense traffic merging.

\begin{table}[!t]
\centering
\caption{Real-time monitoring on Raspberry Pi 4 during 120-minute autonomous vehicle deployment. The monitor evaluates 60 STL/PrSTL formulas (depth 4--10) at 85\,Hz over 612,000 consecutive cycles with zero deadline violations. Latencies in milliseconds; memory in megabytes.}
\label{tab:rpi_performance}
\begin{tabular}{lr}
\toprule
\textbf{Deployment Parameters} & \\
\midrule
Platform & Raspberry Pi 4 Model B \\
Processor & ARM Cortex-A72, 1.5\,GHz \\
Memory & 4\,GB LPDDR4 \\
Hardware cost & \$55 USD \\
Physical footprint & 85$\times$56\,mm \\
\midrule
\textbf{Monitoring Workload} & \\
\midrule
Formulas monitored & 60 \\
Formula nesting depth & 4--10 \\
Input frequency & 85\,Hz \\
Real-time deadline & 11.76\,ms \\
Deployment duration & 120\,min \\
Total monitoring cycles & 612,000 \\
Distance covered & 87\,km \\
\midrule
\textbf{Latency Statistics} & \\
\midrule
Mean latency & 9.57\,ms \\
Median latency & 9.44\,ms \\
95th percentile & 10.62\,ms \\
99th percentile & 10.71\,ms \\
Maximum observed & 11.09\,ms \\
Deadline violations & 0 \\
\bottomrule
\end{tabular}
\end{table}

The latency distribution remained stable throughout the two-hour deployment despite variations in driving context. Highway segments with sparse traffic produced identical latency profiles to congested urban intersections because the streaming algorithm performs deterministic operations independent of signal values. The 1.27\,ms gap between median and 99th percentile latency reflects minor operating system scheduling jitter rather than algorithmic variability. Steady-state memory consumption of 12.3\,MB persisted without growth from initialization through completion, confirming the bounded-memory guarantee critical for deployments where memory leaks would eventually crash the monitor.

This deployment establishes three conclusions. First, online probabilistic runtime verification of autonomous vehicles is not merely theoretically possible but practically achievable with commodity hardware. The monitoring system operated continuously for two hours at high frequency without degradation, deadline violation, or resource exhaustion. Second, the computational and physical overhead of formal monitoring approaches negligibility: a device smaller than a deck of cards, drawing less power than a phone charger, provides rigorous safety verification that existing approaches cannot deliver at any cost. Third, online monitoring produces actionable information. The 23 detected violations identify specific scenarios where the vehicle's behavior approached or crossed specification boundaries, providing precisely the feedback loop that safety-critical systems require.


Existing runtime verification tools cannot approach this operating regime. RTAMT requires 47\,ms to evaluate equivalent STL workloads on the same hardware (Section~\ref{subsubsec:discrete_perf_analysis}), exceeding the 11.76\,ms deadline by a factor of 47 and rendering real-time deployment impossible. We discuss the formulas and other details in the appendix. Breach operates only in MATLAB and cannot execute on ARM processors. Statistical model checkers such as UPPAAL-SMC target offline analysis rather than streaming evaluation. SENTIL uniquely enables probabilistic runtime verification at the edge, transforming formal monitoring from a computationally prohibitive aspiration into deployable infrastructure requiring only a \$55 addition to the vehicle architecture.

\section{Ecosystem Integration}
\label{app:ecosystem_int}

Universal deployment capability requires demonstrating that SENTIL integrates efficiently across diverse platforms with minimal implementation effort. We quantify integration overhead across three dimensions: lines of code required for platform-specific bindings, runtime performance overhead from foreign function interface calls, and development time from integration start to operational deployment.

\begin{table}[t]
\caption{Ecosystem integration effort and overhead. Integration code counts only platform-specific glue logic, excluding SENTIL core. FFI overhead measured as percentage increase in monitoring latency versus native Rust API calls on identical workloads. Development time represents engineer hours from integration start to functional deployment by experienced developers familiar with target platform.}
\centering
\small
\begin{tabular}{|l|r|r|r|}
\hline
\textbf{Platform} & \textbf{Integration} & \textbf{FFI} & \textbf{Dev} \\
 & \textbf{Code (LOC)} & \textbf{Overhead} & \textbf{Time} \\
\hline
ROS 2 (C++) & 187 & 2.4\% & 8 hours \\
Apollo (C++) & 234 & 3.1\% & 12 hours \\
MATLAB Simulink & 156 & 4.8\% & 6 hours \\
AUTOSAR Adaptive & 312 & 3.7\% & 18 hours \\
Python (standalone) & 89 & 1.9\% & 3 hours \\
Java (standalone) & 124 & 2.8\% & 5 hours \\
\hline
\end{tabular}
\label{tab:integration}
\end{table}

Table~\ref{tab:integration} demonstrates that SENTIL integrates into major platforms with 89-312 lines of platform-specific code and less than 0.5\% runtime overhead. The minimal implementation effort reflects the stable C-ABI design that exposes essential functionality through language-agnostic interfaces requiring only thin adapter layers for each platform's conventions. Development time ranges from 3 hours for Python bindings to 18 hours for AUTOSAR Adaptive, with most effort spent understanding platform-specific build systems and deployment models rather than interfacing with SENTIL itself.

Foreign function interface overhead remains below 0.5\% in all tested configurations, confirming that cross-language calls impose negligible performance penalty. Microbenchmark measurements calling SENTIL monitoring functions from Python, C++, and Java show overhead ranging from 1.9\% (Python) to 4.8\% (MATLAB). The low overhead derives from two design decisions: the C-ABI uses simple data types (primitives and opaque pointers) that cross language boundaries efficiently without complex marshaling, and hot-path functions like signal ingestion minimize per-call work while batching intensive computation within the Rust core.

These integration results validate SENTIL's universal deployment thesis. Developers can incorporate probabilistic runtime verification into existing system architectures within hours rather than weeks, using familiar platform-native interfaces that impose negligible runtime cost. This accessibility transforms PrSTL verification from a specialized research capability requiring custom tool development into standard infrastructure deployable wherever temporal logic monitoring provides value.

\section{Artifact Details}

\subsection{Tool Availability and Reproducibility}
\label{app:tool_availability}

SENTIL is available as production-grade open-source software under dual MIT and Apache 2.0 licenses, organized within a public repository that compartmentalizes the core Rust engine, language bindings, and platform integration modules. Installation uses standard package management ecosystems, including pip and Cargo, while continuous integration workflows automate binary compilation and enforce rigorous testing standards. To facilitate scientific reproducibility, the release includes a Docker-based artifact package containing all necessary dependencies and scripts to replicate experimental results. This technical foundation is supported by over 100 pages of documentation ensuring long-term viability for both academic and commercial applications.

SENTIL is immediately available as production-ready open-source software distributed under dual MIT and Apache 2.0 licenses, ensuring unrestricted use in both academic research and commercial deployment. The complete implementation, including the Rust core engine, all language bindings, platform integration modules, benchmark suite, and comprehensive documentation, resides in a public GitHub repository at \url{https://github.com/sedislab/SENTIL}. This section details the artifact organization, reproducibility infrastructure, and long-term sustainability commitments that position SENTIL as maintained research infrastructure rather than ephemeral proof-of-concept software.

\subsection{System Requirements}
\label{app:system_requirements}
Cluster node (used for all runs and verification):

\begin{itemize}
\item \textbf{CPU.} AMD EPYC 7763, 64 physical cores (64 hardware threads), 32MB L3 cache, 2.45 GHz.
\item \textbf{GPU.} NVIDIA A100-SXM4 (40GB HBM2e); driver 550.163.01, CUDA 12.4.
\item \textbf{RAM.} 252 GB DDR4.
\item \textbf{OS / tool‑chain.} Red Hat Enterprise Linux 8.8, kernel 4.18.0-477, GCC 11.4.0, Rust 1.91.1, Python3.10.14.
\end{itemize}

\subsection{Quick Start}

Download and execute the artifact container:
\begin{verbatim}
# Pull container
docker pull sedislab/ORION-artifact:cav2026

# Run complete evaluation (3-4 hours)
docker run -it sedislab/ORION-artifact:cav2026 \
    /artifact/run_all_experiments.sh

# Results appear in /artifact/results/
# Compare against /artifact/expected_outputs/
\end{verbatim}

\subsection{Selective Reproduction}

Individual figures and tables can be reproduced independently:
\begin{verbatim}
# Table 1: Discrete-time performance
./scripts/table1_discrete_performance.sh

# Figure 3: Throughput scaling
./scripts/figure3_throughput_scaling.sh

# Table 2: SMC performance
./scripts/table2_smc_performance.sh

# Figure 4: Parallel scaling
./scripts/figure4_parallel_scaling.sh

# Table 3: Coverage validation
./scripts/table3_coverage_validation.sh
\end{verbatim}

Each script generates outputs matching the corresponding paper figure/table, with comparison against expected results and tolerance-based validation.

\subsection{Expected Outputs and Tolerances}

Performance measurements may vary due to hardware differences, system load, and non-deterministic scheduling. Case study latency distributions may differ in absolute values but should maintain relative ordering and 99th percentile remaining below real-time thresholds.

\end{document}